# Flatten the Li-ion Activation in Perfectly Lattice-matched MXene and 1T-MoS₂ Heterostructures via Chemical Functionalization

*Qiye Guan[a], Hejin Yan[a], Yongqing Cai[a,*]*

*[a] Joint Key Laboratory of the Ministry of Education, Institute of Applied Physics and Materials Engineering, University of Macau, Taipa, Macau, China*

*Corresponding author: yongqingcai@um.edu.mo*

**Keywords**: Li-ion battery, 1T-MoS₂, MXene, chemical functionalization, heterostructure

## Abstract

MXene and its derivatives have attracted considerable attention for potential application in energy storage like batteries and supercapacitors owing to its ultrathin metallic structures. However, the complexity of the ionic and electronic dynamics in MXene based hybrids, which are normally needed for device integration, triggers both challenges and opportunities for its application. In this paper, as a prototype of metallic hybrids of MXene, heterostructures consisting of $Ti_3C_2T_2$ (T= None, O and F atoms) and metallic $MoS_2$ (1T phase) are investigated. Through density functional theory, we investigate the interfacial electronic variation, thermal activation, and anode performance in the lithium-ion battery (LIB) of $Ti_3C_2T_2$/1T-$MoS_2$. We found that different surface atomic groups in MXene can significantly alter the affinity, redox reaction and kinetics of Li atoms in the interface of the $Ti_3C_2T_2$ and $1T-MoS_2$. Through examining the three possible pathways of Li by climbing image-nudged elastic band (CI-NEB) and ab-initio molecular dynamics (AIMD) simulation, the diffusion curve becomes significantly flattened from the naked to O- and F-terminated $Ti_3C_2$ MXene with activation barriers reducing from 0.80 to 0.22 and 0.29 eV, respectively, and room-temperature diffusion coefficients increasing from $1.20\times10^{-6}$ to $2.75\times10^{-6}$, $1.70\times10^{-4}$ $cm^2$ $s^{-1}$, respectively. Half-cell anode computation also shows that $Ti_3C_2T_2(O, F)/1T-MoS_2$ has a higher capacity (298.04 and 293.15 mAh $g^{-1}$) and higher voltage up to 3.14 V than the bare one (81.78 mAh $g^{-1}$ and 0.24 V). The functionalization with O or F eliminates the steric hindrance of Li intercalation by breaking the strong interaction between two layers and provides additional adsorption sites for Li diffusion in the meantime. Our work suggests that surface functional groups play a significant role in $Ti_3C_2T_2/1T-MoS_2$ modification and $Ti_3C_2F_2/1T-MoS_2$ with the high diffusion coefficient and theoretical capacity could be a promising anode material for LIBs.

# 1. Introduction

The energy demand grows faster than ever in the modern world. With the decrease of exhaustible energy on the earth, people turn to pay more attention to renewable energy such as rechargeable batteries, wind power, and solar heat. Among these kinds of green energy, rechargeable batteries or secondary batteries have been prominent due to their portability and a variety of usage scenarios. Therefore, a deep and comprehensive study is required in promoting the performance of these batteries such as lithium-ion batteries (LIBs) which are the most successful systems and widely used right now.[1] Exploring new anode material which has good performance in lithium diffusion and storage has been one of the primary targets.[2-6]

MXene, firstly discovered in 2011, is a class of 2-dimensional carbides and nitrides.[7] MXenes can be expressed in a general formula $M_{n+1}X_nT_x$ (where M is the early transition metal, X is carbon or nitrogen, and $n$ is 1-3, and T is surface termination). Normally, MXenes are produced by selectively etching from the parent compounds MAX, where A is an A-group element, such as Al or Si. Since been discovered, MXenes are proved to have great potential and application in energy storage and catalysis because of their highly active surfaces, good conductivity, abundant functional groups, and facile synthesis.[8-11] For instance, $Ti_3C_2$ has been proved to have good performance as anode materials for LIB with 123.6 mAh $g^{-1}$ at 1C rate.[12] Theoretical study identified a low barrier (0.068-0.070 eV) of Li diffusing above the surface.[13,14] Notably, the multiple possibilities of the surface terminations like F, OH, and O enable a unique tunability of the electronic structures which is lack in most of other 2D materials.[15-17] However, different from other atomically thin 2D materials like graphene, the MXene has a thicker layer and the Li atoms cannot be accommodated within the dense backbone layer consisting of the short M-X bond which is tightly bonded.[14,18-20] Therefore, the Li atoms can only be adsorbed above and across the surface which leads to a relatively small capacity. One effective measure to remedy this is to combine MXene with other materials or molecules for composite anode material.[21-24]

MoS$_2$, a typical material of layered transition-metal dichalcogenide, has been widely applied in LIBs due to its high theoretical capacity (670 mAh g$^{-1}$) and attracted great attention owing to its unique crystal structure. A strong lattice and orbital coupling together with a wealth of polymorphs (semiconducting 2H phase and metallic 1T, 1T' phases) also ensure MoS$_2$ for various applications.[25-35] Different from the semiconducting 2H-MoS$_2$, the 1T phase is less prone to capacity fading.[31,32,36] MXene/2H-MoS$_2$ hybrid has been examined as the anode material of LIBs. Kun Ma et al prepared layered MoS$_2$ into 2D Ti$_3$C$_2$ with the subsequent CTAB intercalation, showed a Li$^+$ storage capacity of 340 mAh g$^{-1}$ even at 20 A g$^{-1}$ beyond 1000 cycles.[37] Guangyuan Du et al also synthesized MoS$_2$/Ti$_3$C$_2$T$_x$ (T= O, F, OH) composite by a simple hydrothermal method, exhibited a reversible capacity of 614.4 mAh g$^{-1}$ at 100 mA g$^{-1}$ after 70 cycles.[38] Zongli Hu et al reported a core-shell structure MoS$_2$/Ti$_3$C$_2$ with low Ti$_3$C$_2$ content (8.87 wt%), delivered a capacity of 706.0 mAh g$^{-1}$ after 1390 cycles at 5 A g$^{-1}$.[39] While previous studies[32-39] demonstrate the encouraging charging/discharging performance of MXene/2H-MoS$_2$ hybrid, there is less study on the synergy between MXene and 1T-MoS$_2$. In addition, we speculate the homogeneous metallic nature across MXene and 1T-MoS$_2$ would allow a fast response of lithium atoms to the external electric field due to the absence of the screening built-in potential unavoidable in MXene/2H-MoS$_2$ and other similar metallic/semiconducting interfaces.

These all motivate us to examine the MXene/1T-MoS$_2$ composite which could take advantage of the facile surface decoration of MXene and the ultrathin metallic nature of 1T-MoS$_2$ for LIBs.

In this work, we construct a MXene/1T-MoS$_2$ heterostructure based on the most popular form of MXene with the Ti$_3$C$_2$ type. We find that the Ti$_3$C$_2$ and 1T-MoS$_2$ have a good lattice registration with a small lattice mismatch. The electronic property and the kinetics of LIB of the hetero-layer have been examined by using density functional theory (DFT). We are particularly interested in the different surface groups of MXene on the energetics and kinetics of Li species. By considering the various surface terminations of MXene, representing as Ti$_3$C$_2$T$_2$ (T= None, O, and F atoms), the role of functional groups in the binding and activating of the Li atoms is examined. Interestingly, we find that different surface terminations of MXene would lead to strikingly different Li absorptions and barriers in the Ti$_3$C$_2$T$_2$/1T-MoS$_2$ heterostructure, and properly surface stoichiometric and functionalization would be critical in applications.

## 2. Results and Discussions

### 2.1. Structural Properties and Stability

First, we start with measuring the structural properties of Ti$_3$C$_2$/1T-MoS$_2$ as well as its functionalized derivatives (Ti$_3$C$_2$O$_2$/1T-MoS$_2$ and Ti$_3$C$_2$F$_2$/1T-MoS$_2$). The hexagonal

Ti$_3$C$_2$T$_2$ layer is built from the Ti$_3$AlC$_2$ phase by striping Al atoms. Two carbon atomic layers interleave into three titanium atomic layers forming a Ti$_3$C$_2$ octahedral. In functionalized structures, fluorine or oxygen atoms coordinate with surface Ti atoms respectively. Notably, there is a small lattice mismatch of around 0.038% between Ti$_3$C$_2$ and 1T-MoS$_2$ with relaxed lattice constants of 6.162 and 6.399 Å. This allows the homogeneous and epitaxial stacking between 1T-MoS$_2$ and Ti$_3$C$_2$ and its derivatives.

Here, 1T phase MoS$_2$ is stacked with Ti$_3$C$_2$T$_2$ and six different stacking structures of Ti$_3$C$_2$T$_2$/1T-MoS$_2$ by taking into account of the relative shift of the bilayer. For bare Ti$_3$C$_2$/1T-MoS$_2$ (**Figure 1a**), the separation between the 1T-MoS$_2$ and Ti$_3$C$_2$ is the smallest one, only 2.375 Å, which means a strong interaction between sulfur atoms of 1T-MoS$_2$ and surface titanium atoms of MXene. This is consistent with the S-Ti bonding evidence shown in the experiment.[18,38] For Ti$_3$C$_2$O$_2$/1T-MoS$_2$, two stacking structures were considered: the interfacial O atoms are right beneath the Mo atoms (type I, Figure 1b) and the S atoms at the top surface (type II, Figure 1c). We found that the type-II surface of Ti$_3$C$_2$O$_2$/1T-MoS$_2$ is around 3.2 eV more energetically favorable than the type-I surface.

For fluorine coordinated Ti$_3$C$_2$F$_2$/1T-MoS$_2$, three stacking possibilities are investigated: the F atoms are located under interfacial S atoms and form F-Ti bonds which are vertically aligned. (type I, Figure 1d), and F atoms are under the S atoms at the top surface (type II, Figure 1e), and F atoms are under the Mo atoms (type III, Figure 1f). The type-III Ti$_3$C$_2$F$_2$/1T-MoS$_2$ is around 4.0 and 6.6 eV lower in energy than that of type-II and type-I heterolayer, respectively. In subsequent sections, only those heterostructures with the lowest energy for each case are discussed.

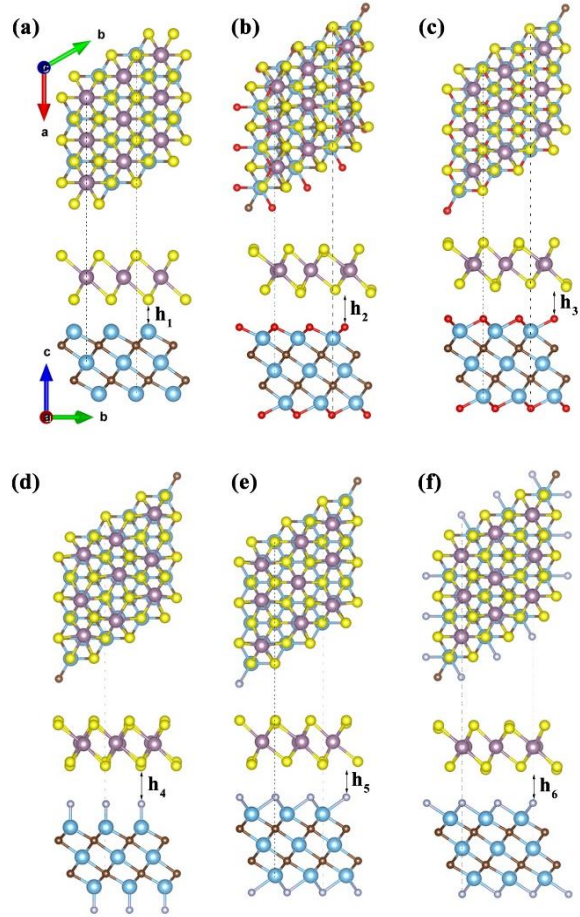

**Figure 1.** Schematic diagram of optimized $Ti_3C_2T_2$/1T-$MoS_2$ structures. (a) Top and side views of $Ti_3C_2$/1T-$MoS_2$, the interlayer distance $h_1$=2.38 Å. (b) and (c) are two shifted forms of $Ti_3C_2O_2$/1T-$MoS_2$, the interlayer distance $h_2$=2.95 Å, $h_3$=2.65 Å. (d), (e) and (f) shows three possible aligned structures of $Ti_3C_2F_2$/1T-$MoS_2$, the interlayer distance $h_4$, $h_5$, $h_6$ are 3.10, 2.78 and 2.72 Å, respectively.

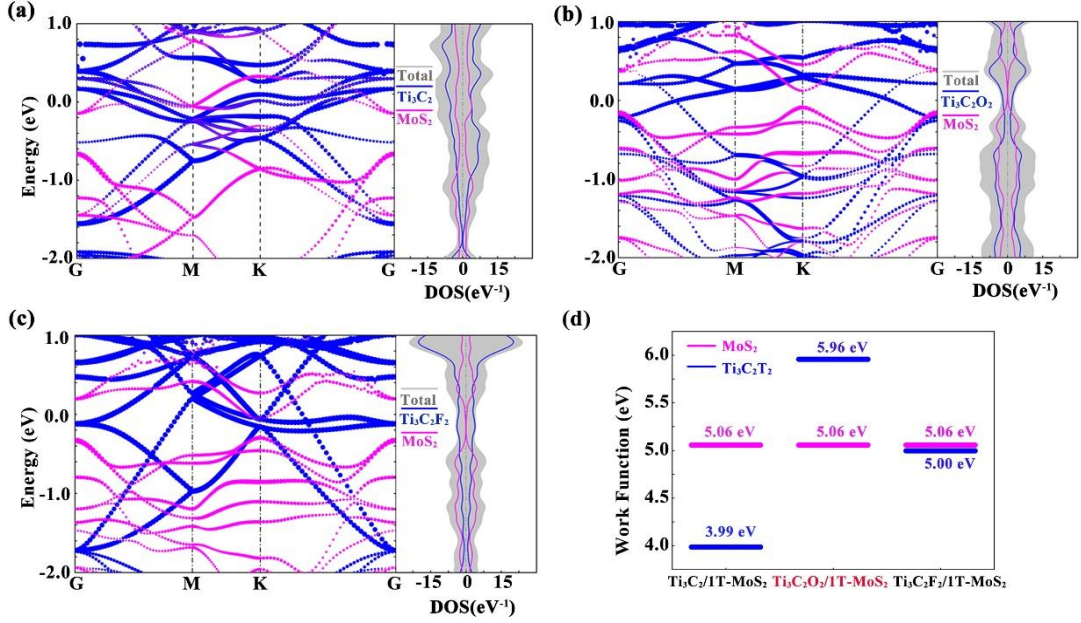

**Figure 2.** Electronic band structures and DOS of (a) $Ti_3C_2$/1T-$MoS_2$, (b) $Ti_3C_2O_2$/1T-$MoS_2$ and (c) $Ti_3C_2F_2$/1T-$MoS_2$, respectively. (d) Comparison of work function of $Ti_3C_2T_2$ and 1T-$MoS_2$.

## 2.2. Electronic Structure

The electronic property of $Ti_3C_2T_x$ is closely related to surface terminations,[15] which could hence affect the electronic structure of the MXene/1T-$MoS_2$ heterostructure. We calculate the band structure for the MXene/1T-$MoS_2$ heterostructures as shown in **Figure 2**. For bare $Ti_3C_2$/1T-$MoS_2$, it exhibits a typical metallic property with the Fermi level ($E_f$) crossing band curves of both the $Ti_3C_2$ and 1T-$MoS_2$ (Figure 2a). The according density of states (DOS) also shows a high peak at the $E_f$, implying a good metallicity. Interestingly, for the O- and F- decorated MXene (Figure 2b and c), the metallicity of $Ti_3C_2O_2$/1T-$MoS_2$ and $Ti_3C_2F_2$/1T-$MoS_2$ is apparently weakened, as indicative of the reduced intensities of the DOS at the $E_f$. Projected band analysis shows that there exists a small band gap opening (~0.21 eV) at K point for 1T-MoS$_2$ in both cases. This is also accompanied with the downward shift of $E_f$ from the case of Ti$_3$C$_2$/1T-MoS$_2$ with some of the MoS$_2$ bands become unoccupied. The band gap opening could be due to the Peierls transition in MoS$_2$ where the electronic energy savings with the band gap around this lower $E_f$ outweighs elastic energy of the disturbed periodic lattice of 1T-MoS$_2$ disturbed by the neighboring F- and O- groups. These results indicate that by tuning surface functional groups on MXene, we can obtain heterostructures with diverse electronic properties. The relative evolution of the $E_f$ level could be explained by the relative alignment of the work function. The work functions of isolated Ti$_3$C$_2$, Ti$_3$C$_2$O$_2$, Ti$_3$C$_2$F$_2$, and 1T-MoS$_2$, are 3.99, 5.96, 5.00 and 5.06 eV, respectively, as shown in Figure 2d. The functionalization of the naked Ti$_3$C$_2$ with the surface anionic –O and –F groups significantly reduces the leakage of the electrons from the interior part which causes the dramatic increase of the work functions of the Ti$_3$C$_2$O$_2$ and Ti$_3$C$_2$F$_2$.

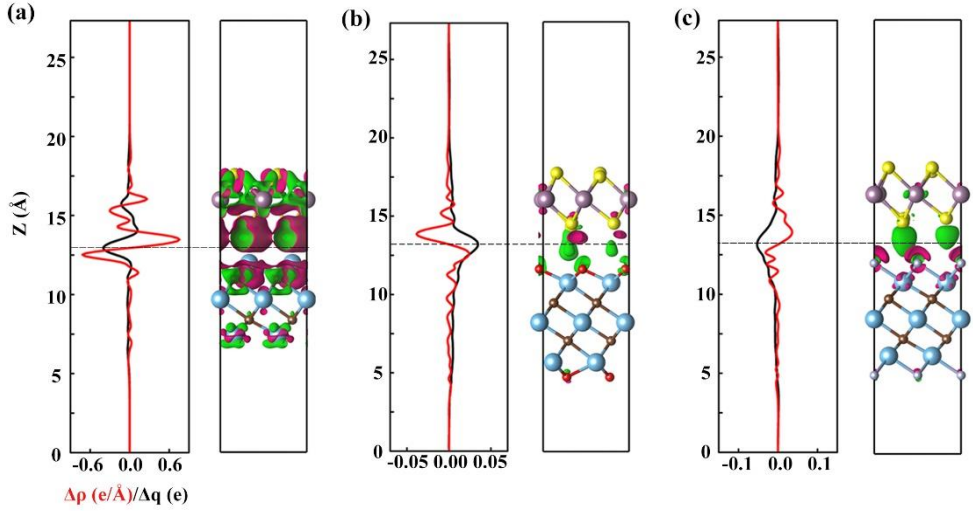

**Figure 3.** Charge transfer across the interlayer of (a) Ti$_3$C$_2$/1T-MoS$_2$, (b) Ti$_3$C$_2$O$_2$/1T-MoS$_2$, (c) Ti$_3$C$_2$F$_2$/1T-MoS$_2$. The amount of transferred charge $\Delta q$ and plane-averaged differential charge density $\Delta \rho$ (along z direction) are represented by black and red, respectively. The red (green) color in the isosurface plots denotes the loss (accumulation) of electrons.

To find out electrons' redistribution after functionalization, we also analyze the charge transfer between the 1T-MoS$_2$ and Ti$_3$C$_2$T$_2$. **Figure 3** shows the isosurfaces of differential charge density (DCD) $\Delta\rho(\mathbf{r})$ and its line-profile curves $\Delta\rho(\mathrm{z})$ along z direction by integrating the in-plane DCD among x-y plane. The total amount of transferred electrons along z direction $\Delta q(z)$ is calculated by the integration of the line-profile curves $\Delta\rho$ (z) from bottom expressed as $\Delta q(z) = \int_{-\infty}^{z} \Delta\rho(z')\,dz'$.[40] Regarding the original Ti$_3$C$_2$/1T-MoS$_2$, the Ti$_3$C$_2$ layer delivers a relatively strong charge transfer of 0.40 electron to MoS$_2$ for the supercell, amount to 0.10 e per S atom. The charge redistribution occurs in the whole 1T-MoS$_2$ sheet, signifying a strong interaction between unpassivated $Ti_3C_2$ and $1T-MoS_2$. The lobes in the DCD isosurface appear at the mid-gap of the interface suggesting a purely ionic interaction between $Ti_3C_2$ and $1T-MoS_2$, largely associated with the interfacial Ti-S interactions. Comparing with the $Ti_3C_2/1T-MoS_2$, O functionalized MXene has a different process as shown in Figure 3b. The trend of charge transfer is opposite with 0.034 electron transferring from $1T-MoS_2$ to $Ti_3C_2O_2$ (-0.017 e per S atom). This implies that the O atoms, each coordinating with two Ti atoms, still are not fully compensated for its $2p^4$ subshell, which triggers more electrons transferred from $1T-MoS_2$. Another reason for this strong charge transfer is due to the much smaller work function of $1T-MoS_2$ than that of the $Ti_3C_2O_2$ as shown above. For the F functionalized $Ti_3C_2F_2/1T-MoS_2$, the trend of electron transfer is the same as the original $Ti_3C_2$ case: 0.053 e (0.013 e per S atom) is transferred from $Ti_3C_2F_2$ to $MoS_2$. Comparing the F- and O- functionalized heterostructure with the pristine MXene case, the much smaller interfacial charge transfer indicates that the $Ti_3C_2$ layer, especially surface Ti atoms are stabilized by functional groups O or F atoms. The insertion of O, F atoms also weakens the bond between S and Ti atoms, which is crucial for the interlayer adsorption and diffusion of alkaline atoms like lithium and sodium. For both decorations, significant amount of electrons are accumulated within the interface between $Ti_3C_2O_2/Ti_3C_2F_2$ and $1T-MoS_2$, reflecting van der Waals plus the slightly covalent nature of the interfacial bonding of S and O/F atoms which is different from the purely ionic interfacial bonding of the $Ti_3C_2/1T-MoS_2$.

## 2.3. Lithium Adsorption and Diffusion

Next, we explore the energetics and kinetics of Li in the interlayer of $Ti_3C_2T_2/1T$-$MoS_2$. First, the interaction between Li atoms and the heterostructure is investigated by comparing the binding energies ($E_b$) for different MXene being involved. The adsorption sites are shown in **Figure 4a**. $E_b$ is calculated according to the equation showing below:

$$E_b = \ E_{total} - E_{Ti_3C_2T_2/1T-MoS_2} - E_{Li}$$

(1)

where $E_{total}$ and $E_{Li}$ are the total energies of Li-$Ti_3C_2T_2/1T$-$MoS_2$ and isolated Li atom, respectively. Original $Ti_3C_2/1T$-$MoS_2$ exhibits a quite small $E_b$ (negative) with -0.24 eV, while O, F functionalized types deliver a significantly promoted binding of -3.14 and -2.36 eV, respectively. This difference reveals that surface functional groups, especially O atoms enhance the performance of Li adsorption. Comparing with the adsorption energy of Li within $Ti_2CO_2$ with $2H$-$MoS_2$ heterostructure (-2.30 eV),[39] functionalized $Ti_3C_2O_2/1T$-$MoS_2$ behaves better. Such a large $E_b$ value ensures that Li would less likely to form clusters[19,41,42] during the diffusion process which will improve the safety as an anode for LIB. In addition, we also computed the adsorption ability of Li on both top and bottom surfaces of this heterostructure. For the top surface, the 1T-MoS2 layer, $Ti_3C_2/1T$-$MoS_2$ exhibits extremely high adsorption energy, while $Ti_3C_2O_2/1T$-$MoS_2$ and $Ti_3C_2F_2/1T$-$MoS_2$ are lower than the prototypes. For the bottom surface, similar results are achieved. Clearly, functional groups' insertion would lead to the increment of the adsorption of Li. Despite this, for better guidance of experiments, we also consider the variance of surface F/O atoms. As shown in table 1, by adjusting the ratio of F and O atoms from 1:3 to 3:1, the shift of adsorption energy is around 0.2 eV which is relatively low that would not influence the total performance seriously.

**Table 1.**

Binding energies and diffusion barriers of MXene/1T-$MoS_2$, * represents the top of $MoS_2$ surface and ** represents the bottom of MXene surface.

| Materials | Binding Energy $E_b$ (eV) | Diffusion Barrier (eV) |
|:---:|:---:|:---:|
| $Ti_3C_2$/1T-$MoS_2$ | -0.24 | 0.80 |
| $Ti_3C_2$/1T-$MoS_2$* | -3.56 | 0.17 |
| $Ti_3C_2$/1T-$MoS_2$** | -2.00 | 0.012 |
| $Ti_3C_2O_2$/1T-$MoS_2$ | -3.14 | 0.22 |
| $Ti_3C_2O_2$/1T-$MoS_2$* | -2.17 | 0.32 |
| $Ti_3C_2O_2$/1T-$MoS_2$** | -2.66 | 0.39 |
| $Ti_3C_2F_2$/1T-$MoS_2$ | -2.36 | 0.29 |
| $Ti_3C_2F_2$/1T-$MoS_2$* | -1.75 | 0.37 |
| $Ti_3C_2F_2$/1T-$MoS_2$** | -1.22 | 0.26 |
| $Ti_3C_2O_1F_1$/1T-$MoS_2$ | -2.50 | - |
| $Ti_3C_2O_{0.5}F_{1.5}$/1T-$MoS_2$ | -2.60 | - |
| $Ti_3C_2O_{1.5}F_{0.5}$/1T-$MoS_2$ | -2.72 | - |

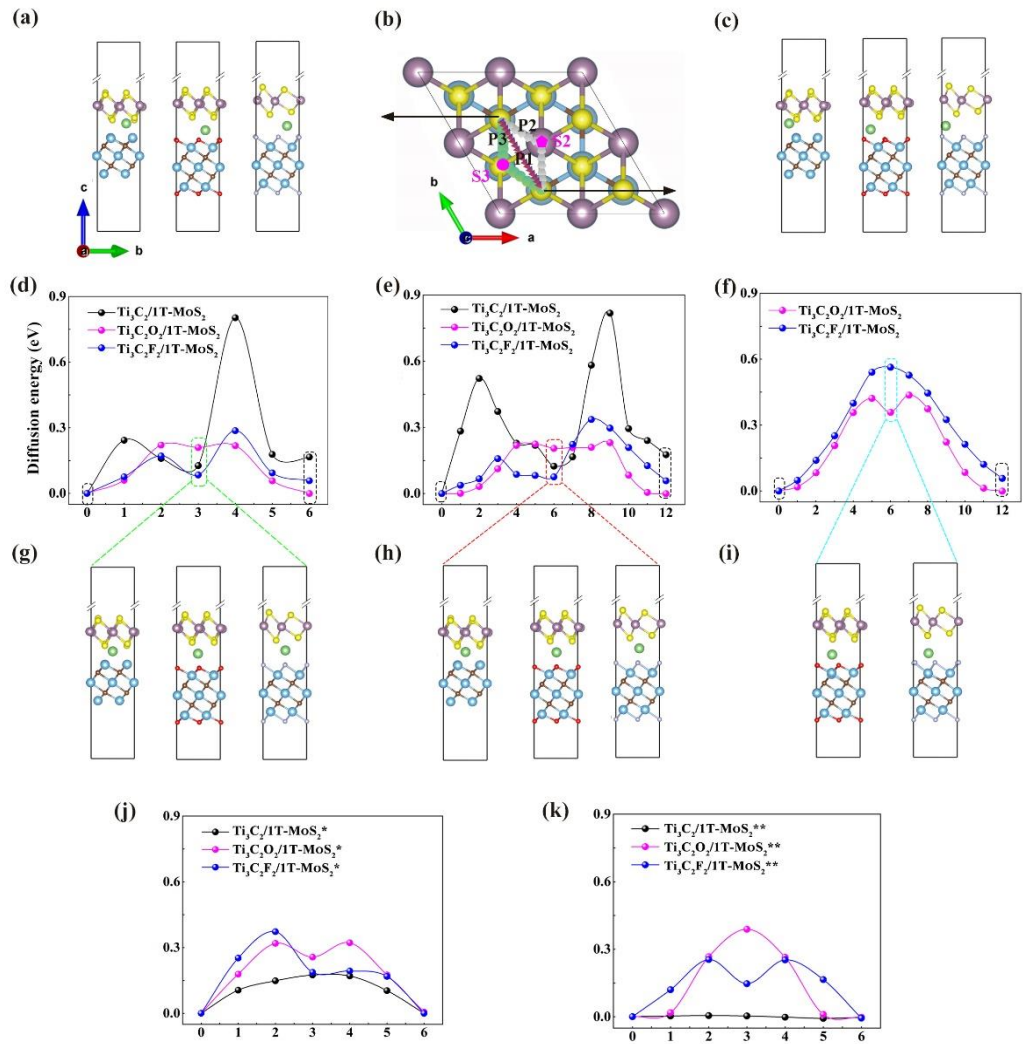

**Figure 4.** Schematic diagram of CI-NEB, (a) side view of initial positions, (b) top view of three possible paths P1, P2, P3 of $Ti_3C_2T_2$/1T-$MoS_2$, (c) side view of final positions. Diffusion barrier through (d) P1, (e) P2 and (f) P3. Optimized structures of energy minimum points in (g) P1, (h) P2 and (i) P3 with an order of $Ti_3C_2$/1T-$MoS_2$ (left panel), $Ti_3C_2O_2$/1T-$MoS_2$ (middle panel), $Ti_3C_2F_2$/1T-$MoS_2$ (right panel). Note for P3 only $Ti_3C_2O_2$/1T-$MoS_2$ and $Ti_3C_2F_2$/1T-$MoS_2$ are considered. Diffusion barrier through (j) top surface and (k) bottom surface.

To further understand the diffusion behavior of Li during the charging and discharging process, the diffusion energy barriers from different migration paths are calculated by the climbing-image nudged elastic band (CI-NEB) method. As shown in Figure 4b, by utilizing the symmetry of the $Ti_3C_2T_2$/1T-$MoS_2$, we design three possible pathways (denoted as P1, P2, and P3) between the two nearest neighboring adsorption sites of Li: initial state (IS) and final state (FS). Figure 4a and c display configurations of IS and FS corresponding to the three different $Ti_3C_2T_2$/1T-$MoS_2$ heterostructures. As shown in Figure 4b, along P1 (IS→FS) path, the Li atom adopts a straightforward pathway. For P2 (IS→S2→FS) path, the Li atom is supposed to migrate from IS to a metastable state S2 (shown in Figure 4b), then from S2 to IS. The Li in S2 structure is placed directly under a Mo atom (the nearest layer to $Ti_3C_2$) while in the middle of three Ti atoms for $Ti_3C_2$ or right above surface O or F atom for $Ti_3C_2T_2$ case. For P3 (IS→S3→FS) route, the Li atom is supposed to diffuse from IS to S3 state (shown in Figure 4b), where the Li is right beneath a S atom and above a Ti atom (the neighboring surface layer of MXene), and next migrate to FS.

First, for both P1 and P2, for all three types of $Ti_3C_2T_2$/1T-$MoS_2$, metastable states are found and shown in Figure 4g and h, respectively. According to Figure 4g, these local energy-minimum points adopt similar structures from P1. Although the $Ti_3C_2$/1T-$MoS_2$ has the weakest $E_b$ comparing with the others, its energy barrier for Li diffusing along P1 is the largest, around 0.80 eV. This is caused by the strong interfacial adhesion between MoS$_2$ and Ti$_3$C$_2$ layer where a narrow interlayer gap of 2.38 Å leads to huge steric hindrance and Coulomb repulsion hence upgrading the migration barrier. For bare Ti$_3$C$_2$/1T-MoS$_2$, the saddle state of P1 corresponds to Li locating in the hollow site of four S atoms. The distance between S and Li is around 2.09 Å which leads to a strong fixation of Li. While for Ti$_3$C$_2$O$_2$/1T-MoS$_2$ and Ti$_3$C$_2$O$_2$/1T-MoS$_2$, the average distance between S-Li is 2.35 and 2.32 Å, and the bond between O-Li is 1.81 Å and F-Li is 1.79 Å. For O and F functionalized heterostructures, the addition of surface atoms not only provides adsorption sites for Li, but also weakens the chemical bonding between layered MoS$_2$ and Ti$_3$C$_2$. Therefore, the diffusion barrier is much lower than that of the original Ti$_3$C$_2$/1T-MoS$_2$ (0.80 eV), which is only 0.22 eV for Ti$_3$C$_2$O$_2$/1T-MoS$_2$ and 0.29 eV for Ti$_3$C$_2$F$_2$/1TMoS$_2$, both are lower than Li diffusing in the single phase of Ti$_3$C$_2$F$_2$ (0.36 eV)[14], in the meantime slightly higher than of pure bilayer MoS$_2$ (0.32 eV)[34]. Overall, the introduction of surface functional groups induces a much lower activation barrier of Li in the interlayer of Ti$_3$C$_2$T$_2$/1T-MoS$_2$.

For P2 pathway (Figure 4h), the diffusion barrier is slightly higher than P1 for all three heterostructures. The barriers are 0.82, 0.23 and 0.34 eV for Ti$_3$C$_2$/1T-MoS$_2$, Ti$_3$C$_2$O$_2$/1T-MoS$_2$ and Ti$_3$C$_2$F$_2$/1T-MoS$_2$, respectively. This difference can be explained by the more rigorous pathway which resulting in higher diffusion energy. For diffusion along P3 pathway (Figure 4i), the bare Ti$_3$C$_2$/1T-MoS$_2$, due to the intact cohesion with a short bonding length of 2.38 Å between S and Ti atoms, the spatial repulsion in S3 is so large that is energetically unreasonable and will lead to structural collapse. Therefore, only $Ti_3C_2O_2$/1T-$MoS_2$ and $Ti_3C_2F_2$/1T-$MoS_2$ are allowed for diffusion along P3 pathway. However, both structures show a higher diffusion energy barrier than P1 and P2, with 0.44 and 0.56 eV for $Ti_3C_2O_2$/1T-$MoS_2$ and $Ti_3C_2F_2$/1T-$MoS_2$ respectively. These increasing barriers are caused by the increased spatial hindrance and van der Waals reaction between S and Ti at S3. The local energy-minimum point for Li diffusion on $Ti_3C_2O_2$/1T-$MoS_2$ is identified to be right beneath the S atom and in the middle of three O atoms, in which the Li atom is located around 2.20 Å between oxygen atoms and 2.30 Å from the S atom. The corresponding energy barrier is 0.44 eV. For $Ti_3C_2F_2$/1T-$MoS_2$, the barrier site also appears when Li is in the hollow site of three F atoms and right beneath the S atom. Li-ion is oriented 2.26 Å from the S atom, and the average distance of Li-F is 2.01 Å. Despite diffusion behavior in the interlayer, Li migration on the top and bottom surfaces along the direct path is also computed as shown in Figure 4j and k. For the top layer, $MoS_2$, all of them exhibit the highest diffusion barrier (0.17~0.37 eV) and the functionalization increases the diffusion barrier by around 0.17 eV. As expected, Li diffusion on the bottom layer ($Ti_3C_2T_2$) is inhibited by the functionalization atoms, especially O (0.39 eV). Comparing Li diffusion ability at different surfaces, the interlayer with a moderate diffusion barrier is the relatively ideal one for electrochemical reactions of LIBs.

From the above discussion, by comparing the calculated diffusion energy barrier along three possible pathways P1, P2, and P3, P1 is obviously most energetically favorable for all three kinds of heterostructures. The diffusion barrier among them follows the sequence of $Ti_3C_2O_2/1T-MoS_2 < Ti_3C_2F_2/1T-MoS_2 < Ti_3C_2/1T-MoS_2$, which means the interlayer functionalization with O and F will prompt the diffusion of Li between the interlayer of $Ti_3C_2T_2$ and $1T-MoS_2$.

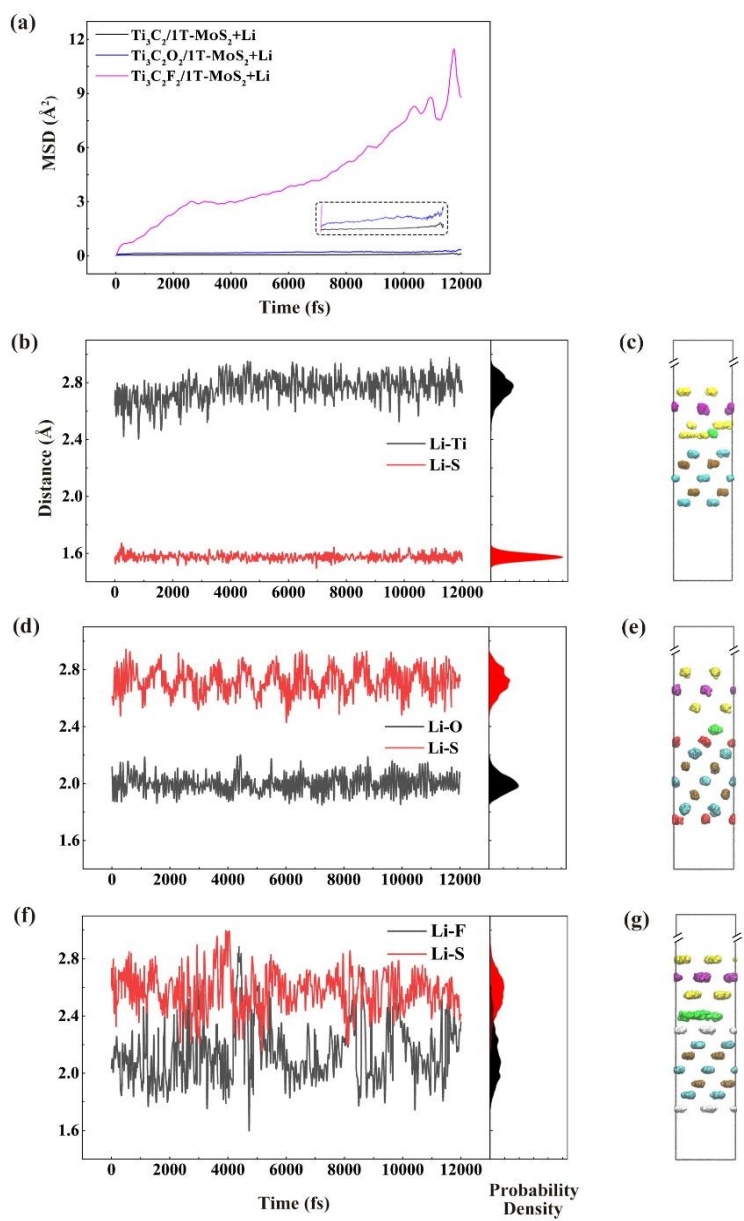

**Figure 5.** Mean square displacements (MSDs) of Li atom at (a) 300 K. The inset shows the enlarged curve of $Ti_3C_2/1T-MoS_2$ and $Ti_3C_2O_2/1T-MoS_2$. Variation of atom

distances in (b) $Ti_3C_2$/1T-$MoS_2$, (d) $Ti_3C_2O_2$/1T-$MoS_2$, and (f) $Ti_3C_2F_2$/1T-$MoS_2$ during molecular dynamics simulations. Multi-frame images of trajectories at 300K of (c) $Ti_3C_2$/1T-$MoS_2$, (e) $Ti_3C_2O_2$/1T-$MoS_2$, and (g) $Ti_3C_2F_2$/1T-$MoS_2$, respectively.

## 2.4. Ab Initio Molecular Dynamics

We further performed ab initio molecular dynamics (AIMD) to investigate the dynamic behavior of the Li atom migrating across the interlayer between MXene and 1T-$MoS_2$ (**Figure 5**). We are particularly interested in the effect of the surface-functionalized groups on the migrating behaviors of Li atom. To this end, here we analyze the mean square displacements (MSDs) from which the diffusion coefficient ($D$) which dominates the charge and discharge rate is also computed which is shown below:

$$D = \frac{MSD(t)}{2d\Delta t} = \frac{\langle |x_i(t)-x_i(t_0)|^2 + |y_i(t)-y_i(t_0)|^2 + |z_i(t)-z_i(t_0)|^2 \rangle}{2d\Delta t} \qquad (2)$$

where $x_i$, $y_i$ and $z_i$ are coordinates of lithium atom $i$. $d$ is the dimension of the structure, $\Delta t$ is the time relative to the initial moment. For $Ti_3C_2T_2$/1T-$MoS_2$, $d$ is taken as 2, and $\Delta t$ is taken between 1 ps to 12 ps to eliminate the nonlinear snippet.[43] First of all, we examine the AIMD at 300 K. The initial position of the Li atom is placed at the adsorption site with energetical preference. As can be seen in Figure 5a, $Ti_3C_2F_2$/1T-$MoS_2$ exhibits a much higher value of MSD. The corresponding trajectory (Figure 5g) also identifies the highly "active" nature of the Li atom between F-decorated MXene and 1T-MoS$_2$. According to our previous results, the Ti$_3$C$_2$/1T-MoS$_2$ has the lowest $E_b$ (-0.24 eV) which means an easier migration of Li. Nevertheless, the relatively narrow gap separating the MoS$_2$ and Ti$_3$C$_2$ leads to a spatial hindrance and restricts the migration of Li. This is reasonable and consistent with the large activation barrier (0.80 eV) from our CI-NEB calculation. For Ti$_3$C$_2$O$_2$/1T-MoS$_2$, the Li also has a relatively limited diffusion rate which on the other hand could be due to the largest $E_b$ (-3.14 eV) with Li albeit with a moderate diffusion barrier (0.22 eV). In contrast, Ti$_3$C$_2$F$_2$/1T-MoS$_2$ is most suitable for Li diffusion owing to the co-existence of a moderate $E_b$ (-2.36 eV) and a small barrier (0.29 eV). The derived $D$ of Ti$_3$C$_2$/1T-MoS$_2$, Ti$_3$C$_2$O$_2$/1T-MoS$_2$, and Ti$_3$C$_2$F$_2$/1T-MoS$_2$ are 1.20×10$^{-6}$, 2.75×10$^{-6}$, and 1.70×10$^{-4}$ cm$^2$ s$^{-1}$, respectively. Comparing with the diffusion coefficient of Li in graphite: C$_{12}$Li (1.46×10$^{-11}$ cm$^2$ s$^{-1}$),[44, 45] the Ti$_3$C$_2$T$_2$/1T-MoS$_2$ exhibits an inherent much higher activity.

To further exploring the role of functional group in Li diffusion, the variation of bond distance of Li-S, Li-F and Li-O are calculated as shown in Figure 5b, d, and f. The probability density is given by the atom distance distribution with a bin size of 0.02 Å. Bare Ti$_3$C$_2$/1T-MoS$_2$ exhibits a strong binding between Li and S atoms with an average 1.57 Å bond length, which leads to a distortion of the nearby S atom as can be seen in Figure 5c. In contrast, Ti$_3$C$_2$O$_2$/1T-MoS$_2$ and Ti$_3$C$_2$F$_2$/1T-MoS$_2$ show a stable MoS$_2$ layer during AIMD with 2.72 and 2.57 Å Li-S bond due to the wider interlayer gap and the stronger adsorption ability of F, O atoms. Comparing the Li-T (O, F atom) distance between Ti$_3$C$_2$O$_2$/1T-MoS$_2$ (2.00 Å) and Ti$_3$C$_2$F$_2$/1T-MoS$_2$ (2.14 Å), with a smaller average bond length and centralized distribution of probability density, O atoms exhibit a stronger fixation capacity than F atoms, leading to a sluggish Li diffusion behavior. The wide distribution of Li-F and Li-S bond length in $Ti_3C_2F_2$/1T-$MoS_2$ also identifies the active diffusion behavior of Li as shown in Figure 5f and g.

As discussed above, it is evident that functional groups of MXene affect the thermal activity of intercalated species, and according to our results, the F functionalization is most favorable for the fast diffusion of the Li atom.

## 2.5. Theoretical Li Storage Capacity and Open-Circuit Voltage

In this part, we investigate the performance of $Ti_3C_2T_2$/1T-$MoS_2$ as anode materials for LIBs through the calculation of theoretical Li storage capacity and open-circuit voltage (OCV) with different loads of Li. As shown in **Figure 6a** and **b**, the max capacity of Li atoms is found by filling all energetically favorable sites: A, B, C, and D. For bare $Ti_3C_2$/1T-$MoS_2$, only one site is eligible for Li adsorption. The theoretical capacity can be calculated through the following equation:[13]

$$C = \frac{xnF}{M} \tag{3}$$

where $x$ is the number of absorbed atoms, $n$ is the valence number of Li, $F$ is the Faraday constant (26801 mAh $g^{-1}$), and $M$ is the atomic mass of $Ti_3C_2T_2$/1T-$MoS_2$ (327.7, 359.7, and 365.7 g $mol^{-1}$ for $Ti_3C_2$/1T-$MoS_2$, $Ti_3C_2O_2$/1T-$MoS_2$, and $Ti_3C_2F_2$/1T-$MoS_2$, respectively). Based on a full load of Li atoms at energetically favorable sites and the radius of Li of 0.76 Å, the max theoretical capacity of $Ti_3C_2$/1T-$MoS_2$, $Ti_3C_2O_2$/1T-MoS$_2$, and Ti$_3$C$_2$F$_2$/1T-MoS$_2$ are 81.78, 298.04, and 293.15 mAh g$^{-1}$, respectively. Compared with the reported single Ti$_3$C$_2$F$_2$ (130.0 mAh g$^{-1}$),[14,15] the Ti$_3$C$_2$F$_2$/1T-MoS$_2$ has a noticeable improvement, which is comparable to graphite. While it is still lower than pure 1T-MoS$_2$ (670.0 mAh g$^{-1}$) as anode materials for LIBs.[32-35]

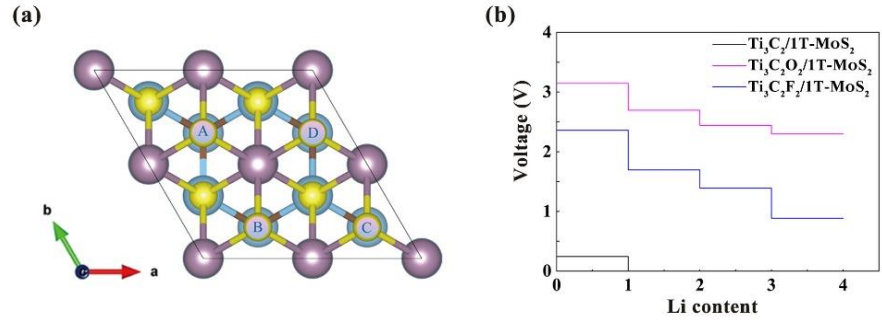

**Figure 6.** (a) Adsorption sites of Ti$_3$C$_2$T$_2$/1T-MoS$_2$ (denoted as A, B, C, and D). (b) OCV of different Li contents in Ti$_3$C$_2$T$_2$/1T-MoS$_2$.

The OCV is another critical metric for the anode material of LIBs. The OCV can be computed by the following equation:[46, 47]

$$V = -\frac{\Delta G}{nF} \approx -\frac{\Delta E}{nF} \qquad (4)$$

where $\Delta G = \Delta E + P\Delta V - T\Delta S$ is the difference in Gibbs energy. Here $P\Delta V$ can be neglected ($10^{-5}$ eV) and $T$ is set to 0 K, $\Delta G$ could be mainly dominated by the variation of the total energy $\Delta E$. The OCV obtained from $\Delta E$ can be calculated as:

$$V \approx \frac{E_{C_1} - [E_{C_2} + (C_1 - C_2)E_{Li}]}{(C_1 - C_2)e^-} \qquad (5)$$

where $E_{C_1}$ and $E_{C_2}$ are total energies of two heterostructures with Li concentrations $C_1$ and $C_2$ ($C_1 > C_2$). As shown in Figure 6b, with the increment of adsorbed Li atoms, the corresponding OCV gradually reduced. For Ti$_3$C$_2$/1T-MoS$_2$, the OCV is 0.24 V. For the Ti$_3$C$_2$O$_2$/1T-MoS$_2$, the initial OCV is 3.14 V, after insertion more Li, the corresponding OCV gradually reduced to 2.30 V. For Ti$_3$C$_2$F$_2$/1T-MoS$_2$, similar to Ti$_3$C$_2$O$_2$/1T-MoS$_2$, with the uptake of more Li atoms, the corresponding OCV gradually decreases from 2.36 to 0.88 V. Regarding practical application, when the operating voltage of the anode is lower than 1 V versus pure Li, the Fermi energy of the anode will be lower than the lowest unoccupied molecular orbital (LUMO) of the organic electrolytes, resulting in the decomposition of solid electrolyte interphase (SEI) and electrolyte.[48] Therefore, a higher OCV associated with decorating functional groups F and O atoms will increase the performance of MXene/MoS$_2$ as anode materials for LIB. In addition, a higher OCV also is beneficial for avoiding Li plating and a high rate capability.[48]

## 3. Conclusions

In summary, via first-principles calculations, we explored the interfacial electronic properties of heterostructures consisting of MXene and 1T-MoS$_2$ which have a small lattice mismatch allowing homogeneous and epitaxial integration. We found that the surface functional groups in MXene can dramatically alter the interfacial electronic affinity, that the –O and –F terminated surface will significantly increase the work function of MXene and thus reducing the leakage of the electrons. By adjusting surface functional groups (-O, -F), the total performance of $Ti_3C_2T_2$/1T-$MoS_2$ is greatly promoted. The insertion of F, O atoms leads to a smaller charge transfer and weakening of S-Ti bonds, which is necessary for the uptake (i.e. stronger adsorption) and activation (i.e. more mobile) of Li atoms. Our AIMD results also confirm the fast kinetics of Li atoms in residual-contained MXene heterostructure. In particular, we reveal that O atoms exhibit stronger fixation capacity than F atoms, leading to a sluggish Li diffusion behavior. The high diffusivity and great capacity suggest that MXene derivatives and hybrids are highly promising for LIBs applications.

# 4. COMPUTATIONAL METHODS

First-principles calculations are carried out by the plane wave code Vienna ab initio simulation package (VASP)[49] with projector augmented wave (PAW) method. Exchange-correlation energy is performed under the generalized gradient approximation (GGA) with the form of Perdew-Burke-Ernzerhof (PBE).[50] DFT-D3 functional with Grimme[51] correction is employed to describe the weak van der Waals (vdW) interaction between 1T-$MoS_2$ and MXenes. An energy cutoff of 450 eV and energy convergence of $10^{-4}$ eV are used in all calculations. All of the structures are optimized until the forces exerted on each atom are <0.005 eV Å$^{-1}$. A 2×2×1 supercell is adopted and the Brillouin zone K-point mesh is set as 4 × 4 × 1 for structural optimization, then a 12 ×12 ×1 K-point mesh is applied for electronic structure computations. The thickness of the vacuum region is set to >15 Å to avoid interference of periodic images. The diffusion barriers are calculated through the climbing-image nudged elastic band (CI-NEB).[52] The ab initio molecular dynamics simulations in the NVT ensemble are employed based on the Nosé-Hoover thermostat[53] with a time step of 1fs.

## Acknowledgments

This work is supported by the University of Macau (SRG2019-00179-IAPME) and the Science and Technology Development Fund from Macau SAR (FDCT-0163/2019/A3), the Natural Science Foundation of China (Grant 22022309) and Natural Science Foundation of Guangdong Province, China (2021A1515010024). This work was performed in part at the High Performance Computing Cluster (HPCC) which is supported by Information and Communication Technology Office (ICTO) of the University of Macau.